%% file: main.tex
\documentclass[sigconf, nonacm]{acmart}

\usepackage[T1]{fontenc}

\usepackage{subcaption}

\usepackage{tikz}
\usepackage{amsmath}
\usepackage{siunitx}
\usepackage{booktabs}
\usepackage{multirow}
\usepackage{enumitem}
\usepackage{mathtools}
\usepackage{balance}
\usepackage{pifont}
\usepackage[normalem]{ulem}
\usepackage[toc,page]{appendix}

\copyrightyear{2022} 
\acmYear{2022} 
\setcopyright{acmcopyright}\acmConference[ASIA CCS '22]{Proceedings of the 2022 ACM Asia Conference on Computer and Communications Security}{May 30-June 3, 2022}{Nagasaki, Japan}
\acmBooktitle{Proceedings of the 2022 ACM Asia Conference on Computer and Communications Security (ASIA CCS '22), May 30-June 3, 2022, Nagasaki, Japan}
\acmPrice{15.00}
\acmDOI{10.1145/3488932.3497771}
\acmISBN{978-1-4503-9140-5/22/05}

\usepackage[absolute]{textpos}
\begin{document}

\begin{textblock}{19}(1.5,2)
\noindent\large This paper was accepted for publication at the ACM ASIA Conference on Computer and Communications Security (AsiaCCS) 2022.
\end{textblock}

%
\title{Signal Injection Attacks against CCD Image Sensors}

\author{Sebastian Köhler}
\affiliation{%
   \institution{University of Oxford}
   \country{United Kingdom}}
\email{sebastian.kohler@cs.ox.ac.uk}

\author{Richard Baker}
\affiliation{%
   \institution{University of Oxford}
   \country{United Kingdom}}
\email{richard.baker@cs.ox.ac.uk}

\author{Ivan Martinovic}
\affiliation{%
   \institution{University of Oxford}
   \country{United Kingdom}}
\email{ivan.martinovic@cs.ox.ac.uk}

\input{00-Abstract/abstract}

\maketitle

\input{01-Introduction/introduction}

\input{02-RelatedWork/relatedwork}

\input{03-Background/background}

\input{04-ThreatModel/threat_model}

\input{05-Attack/attack}

\input{06-Evaluation/evaluation}

\input{07-Limitations/limitations}

\input{08-Countermeasures/countermeasures}

\input{09-Conclusion/conclusion}

\input{10-Acknowledgement/acknowledgement}

\bibliographystyle{ACM-Reference-Format}
\bibliography{references}

\balance

\clearpage

\input{11-Appendix/appendix}

\end{document}

%% file: 00-Abstract/abstract.tex
\begin{abstract}

Since cameras have become a crucial part in many safety-critical systems and applications, such as autonomous vehicles and surveillance, a large body of academic and non-academic work has shown attacks against their main component --- the image sensor.
However, these attacks are limited to coarse-grained and often suspicious injections because light is used as an attack vector.
Furthermore, due to the nature of optical attacks, they require the line-of-sight between the adversary and the target camera.

In this paper, we present a novel post-transducer signal injection attack against CCD image sensors, as they are used in professional, scientific, and even military settings.
We show how electromagnetic emanation can be used to manipulate the image information captured by a CCD image sensor with the granularity down to the brightness of individual pixels.
We study the feasibility of our attack and then demonstrate its effects in the scenario of automatic barcode scanning.
Our results indicate that the injected distortion can disrupt automated vision-based intelligent systems.

\end{abstract}

%% file: 01-Introduction/introduction.tex
\section{Introduction}

Over the last few decades, the underlying architecture of image sensors has experienced a significant shift in technology. 
Nowadays, two major image sensor architectures exist --- Complementary Metal-Oxide-Semiconductor (CMOS) and Charge-Coupled Device (CCD) image sensors.
Due to the improved semiconductor manufacturing process, the production costs of CMOS image sensors have decreased immensely, while the performance of the sensors increased. 
As a result, CMOS image sensors have almost entirely replaced CCD image sensors in consumer devices, such as mobile and IoT devices, autonomous vehicles, retail, and surveillance.

However, due to their excellent photometric performance and their capability to capture frames without geometric distortions, CCD image sensors are still used in specific professional and scientific applications~\cite{durini-2019, ccd_future_bright}.
The fields of application range from ground and space astronomy~\cite{howell2006handbook, duriscoe2007measuring} over microscopy~\cite{jerome2017practical}, industrial automation~\cite{dfm25g445} to military surveillance and defense systems~\cite{skuljan2017quadcam, hagt2009china}.
With the increasing usage of intelligent systems that make safety-critical decisions based on the trusted captured image information, the integrity of the camera inputs has become crucial.
Various attacks against camera-based systems compromising the integrity have been demonstrated in academic literature~\cite{Kohler2021TheySM, petit2015remote, yan2016can, sayles2020invisible}.
Since image sensors are optical sensors, the most obvious attack vector is the injection of light.
However, injecting light in a controlled way is almost infeasible and only partially possible for CMOS image sensors that implement an electronic rolling shutter mechanism that reads the captured image information row by row, rather than all-at-once (global shutter)~\cite{Kohler2021TheySM, sayles2020invisible}.
In contrast, CCD image sensors always implement a global shutter inherent to their design.
This means fine-grained signal injection attacks using light are not possible.
Moreover, a light-based attack requires line-of-sight between the adversary and the target camera.
Finally, attacks that leverage optical emission tend to be suspicious and easily detected by simple mechanisms.
For example, if a frame is suddenly over- or under-exposed, an alarm is triggered~\cite{synology, bosch-2016-tamperdetection}.

In this paper, we overcome these limitations by using intentional electromagnetic interference (EMI).
We show that fine-grained perturbations can be injected into CCD image sensors using electromagnetic emanation.
While the susceptibility of CCD image sensors against electromagnetic interference 
has been evaluated in the context of electromagnetic compatibility (EMC)~\cite{wacholc2019investigation}, to the best of our knowledge, no research has been conducted from the perspective of an adversary trying to inject fine-grained, controlled perturbations using intentional EMI.
Yet, we demonstrate that, due to their architecture, CCD image sensors are vulnerable to post-transducer signal injection attacks using electromagnetic waves. 

With ideal conditions and information, an attacker could exploit the vulnerability to reproduce arbitrary patterns within the output of the image sensor; however, such conditions are unlikely in the real world. Instead, we show the impact of the attack in a far more achievable setting by disrupting the correct operation of barcode reading, as used heavily in manufacturing and logistics~\cite{weng2012design,sick-autoident-2014}. Such an attack on automated barcode reading is simple to mount but has an immediate economic impact on the victim. 
\\\\
\noindent\textbf{Contributions}
Specifically, we make the following contributions:
\begin{itemize} \itemsep0pt
	\item We present a novel, post-transducer signal injection attack against CCD image sensors and demonstrate how an adversary can gain fine-grained control over the brightness intensity down to individual pixels.
	\item We analyze the susceptibility of two CMOS image sensors against the same attack to underpin our hypothesis that the signal injection attack is possible due to the architecture of CCD image sensors.
	\item We demonstrate the consequences of a signal injection attack against CCD image sensors in the context of automatic barcode scanning as it is heavily used in manufacturing and logistics.
	\item We lay the basis for further evaluation of signal injection attacks against CCD image sensors.
\end{itemize}

%% file: 02-RelatedWork/relatedwork.tex
\section{Related Work}

Academic literature has presented signal injection attacks against a variety of sensors and devices, such as medical devices~\cite{kune2013ghost, rasmussen2009proximity}, voice-controlled personal assistants~\cite{yan2019feasibility, Sugawara2019}, thermometers~\cite{yan-2019}, MEMS inertial sensors~\cite{trippel2017walnut, tu2018injected}, air-pressure sensors~\cite{tu2021transductionasiaccs}, and Advanced Driver Assistance Systems (ADAS)~\cite{yan2016can, Cao2019, xu2018analyzing, man2020ghostimage}.
Depending on the target, the attack vector can range from acoustic waves over optical emission to electromagnetic emanation~\cite{giechaskiel2019sok, yan2020sok}.
Furthermore, signal injection attacks can be differentiated based on the component they are targeting. 
If an untrustworthy sensor measurement is directly injected into the transducer using the same physical quantity the sensor is intended to sense, it is called a pre-transducer attack.
In contrast, in post-transducer signal injection attacks, the signal is induced in any component after the sensing part, for example, into a wire connecting the transducer and the microcontroller, via electromagnetic coupling~\cite{yan2020sok}.
For cameras, which have become a popular target due to their widespread use, pre-transducer attacks using optical radiation as the attack vector are the most obvious route to go.
For instance, shining a laser at the camera of a vehicle is a cheap and effective attack to render its ADAS useless~\cite{petit2015remote, yan2016can}.
With a little more effort, the rolling shutter mechanism in CMOS image sensors can be exploited to execute a more controlled signal injection attack~\cite{Kohler2021TheySM, sayles2020invisible}. 
Although exploiting the rolling shutter is less disruptive than fully blinding the camera, using visible light will always be suspicious and, potentially, be easily detected~\cite{bosch-2016-tamperdetection}.
Moreover, the attack is bound to row-wise injections and therefore only allows coarse perturbations~\cite{Kohler2021TheySM, sayles2020invisible}.
In addition, signal injection attacks involving optical emission require line-of-sight between the adversary and the target camera.
In contrast, leveraging electromagnetic waves as the attack vector gives the adversary precise control over the perturbation from outside the line-of-sight.
In fact, with the appropriate equipment, manipulating the signal charge of individual pixels is possible. 
In this paper, we present a novel post-transducer signal injection attack that enables an adversary to obtain such a capability by exploiting the architectural structures of CCD image sensors using an off-the-shelf software-defined radio.

%% file: 03-Background/background.tex
\begin{figure}[t]
	\centering
	\subcaptionbox*{} {
		\includegraphics[width=0.98\linewidth]{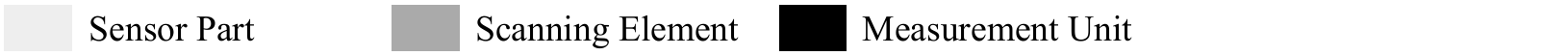}\vspace{-4mm}%
	} 
	\begin{subfigure}[b]{.49\linewidth}
		\centering
		\includegraphics[width=.95\textwidth]{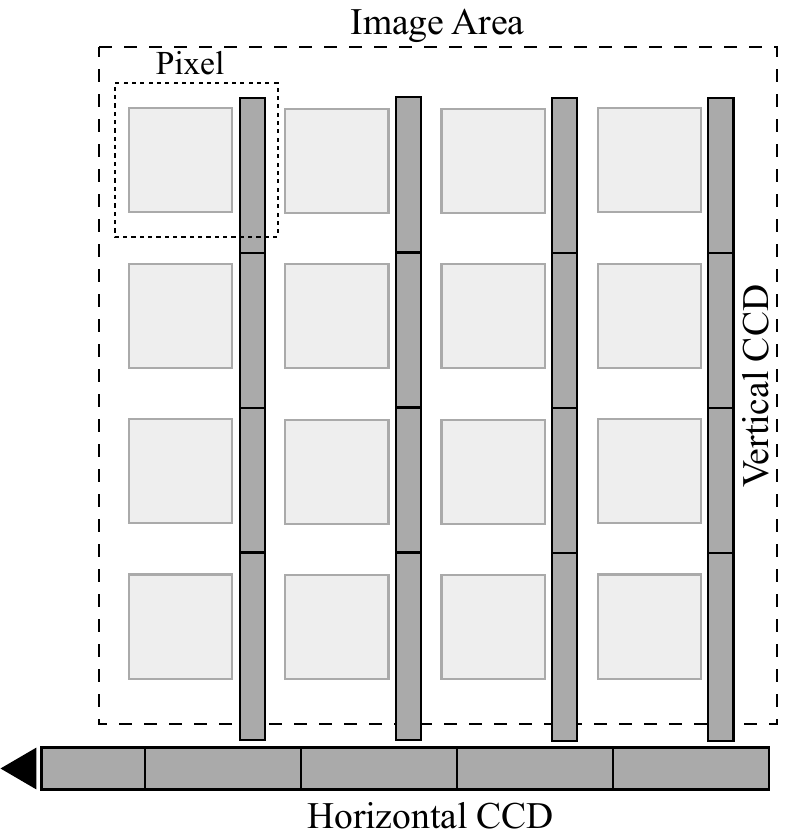}
		\caption{IT-CCD image sensor layout}
		\label{fig:ccd-layout} 
	\end{subfigure}%
	\begin{subfigure}[b]{.49\linewidth}
		\centering
		\includegraphics[width=.95\textwidth]{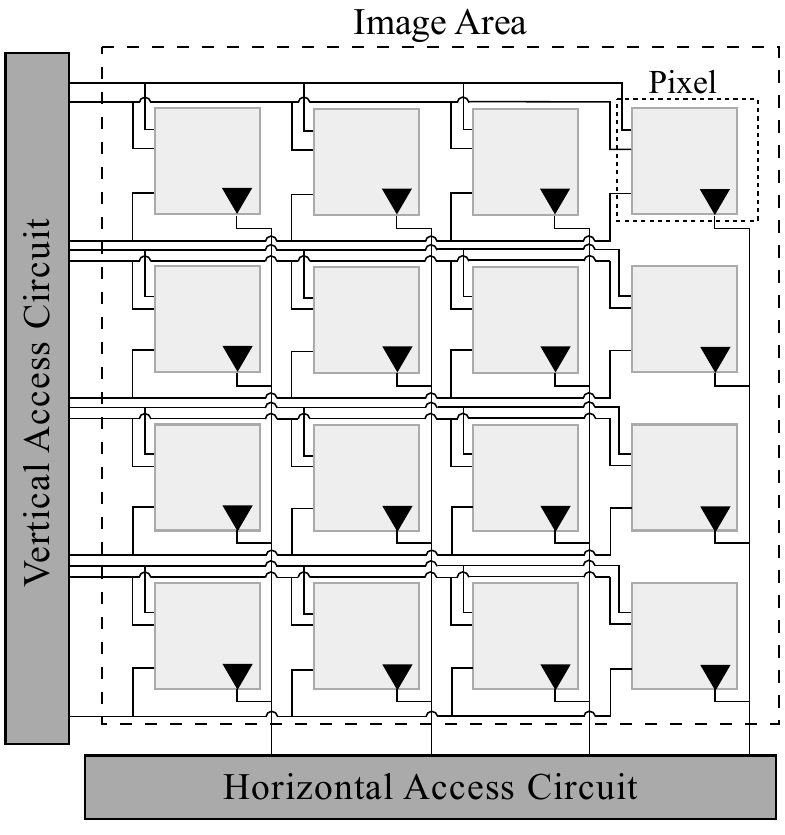}
		\caption{CMOS image sensor layout}
		\label{fig:cmos-layout} 
	\end{subfigure}%
	\caption{Simplified schematic representation of an Interline-Transfer-CCD (IT-CCD) and a CMOS image sensor. The CCD image sensor uses one measurement unit, while the CMOS image sensor implements one per pixel.}
	\label{fig:image-sensor-comparison}
\end{figure}

\section{Image Sensor Fundamentals}\label{sec:background}

Independent of the image sensor architecture, i.e., CMOS or CCD, the fundamental components of an image sensor are the same.
All image sensors have a sensing part that captures the incident light, a scanning element that is responsible for the recovery of the generated signal charge, and a measurement unit that quantifies and amplifies the signal charge.
The main difference between CMOS and CCD image sensors is the order in which these components are arranged. 
While CMOS image sensors have a measurement unit integrated into each pixel, CCD image sensors rely on a single measurement unit~\cite{nakamura-2006}.
A comparison of the two image sensor architectures is depicted in Figure~\ref{fig:image-sensor-comparison}.
In the following, we will focus on the architecture of CCD image sensors. 

\subsection{Photodiode Array}\label{sec:photodiode_array}

The photodiode array is the sensing part of an image sensor.
A two-dimensional array composed of photodiodes, also known as pixels, captures image information in the form of light.
More precisely, incident photons are captured and converted into a signal charge.
The longer a photodiode is exposed to light, the more photons are captured and the higher is the resulting signal charge.
However, each photodiode only captures the intensity for one of the three color channels --- red (R), green (G), or blue (B).
This is achieved by overlaying a Color-Filter Array (CFA) on top of the photodiode array.
The most well-known and most commonly used CFA is the Bayer-Matrix.
Since the human eye perceives green tones more intensively~\cite{nakamura-2006}, the Bayer-Matrix divides the image area into 50\% green, 25\% blue, and 25\% red pixels~\cite{bayer1976colour}. 
To reconstruct an image from the raw color information per pixel, a process known as de-mosaicing is necessary.

Usually, the number of physical pixels (photodiodes) exceeds the maximum resolution of the captured frames.
The additional pixels do not directly contribute to the final images.
However, they provide useful supplementary information, such as color information, and help to determine the boundaries of the frames.

As described earlier, the longer the photodiodes are exposed to incident light, the brighter the resulting image.
To capture enough signal charge in low light conditions, the auto-exposure mechanism of the camera adjusts the exposure time to an optimal value.
Once sufficient signal charge is accumulated, i.e., the integration period finished, the signal charge is read out by the scanning element and transferred to the measurement unit, which will be discussed next.

\subsection{Scanning Element}

The scanning element is responsible for recovering the signal charge from the photodiodes and the transmission to the measurement unit.
In an Interline-Transfer-CCD (IT-CCD) image sensor, the scanning element is composed of multiple shift registers, which are arranged horizontally and vertically and thus are often referred to as H-CCD and V-CCD.
A simplified schematic representation of such an image sensor is depicted in Figure~\ref{fig:ccd-layout}.
Once the integration period for a frame is completed, the generated signal charge is shifted from the photodiodes into the V-CCDs.
For this reason, the horizontal and vertical CCDs can also be seen as a memory buffer~\cite{nakamura-2006}. 
With the shift of the signal charge into the V-CCD, the new integration period starts. 
While the new frame is captured, the signal charge is simultaneously shifted row by row into the H-CCD, before it is measured and amplified by the measurement unit. 
Figure~\ref{fig:ccd_readout} illustrates the readout process of an Interline-Transfer-CCD.
Although CCD image sensors implement a global shutter --- meaning the exposure and signal recovery happens all-at-once --- the digitization is still a sequential process.
This means that the analog-to-digital converter (ADC) samples the pixels one-by-one, starting with the pixel at (0,0).

\subsection{Measurement Unit}\label{sec:measurement_unit}

As the name indicates, the measurement unit is responsible for quantizing and amplifying the captured signal charge per pixel.
The measurement unit usually consists of an ADC and an amplifier.
The ADC samples the analog signal, in the case of the image sensor the signal charge of each pixel, and maps it to a discrete value, usually ranging between 0 and 255.
The exact range depends on the resolution of the ADC.
A higher resolution means that the analog signal can be mapped to more discrete values.
The higher the amplitude of the continuous signal, the higher the discrete value the sample it is mapped to.
Intuitively, higher distinct values represent a higher brightness.
As described before, in poor ambient light conditions, the exposure time has to be extended to capture enough light.
However, depending on the purpose of application, it might not be possible to increase the exposure time further.
Once the integration time is longer than $1/F$ seconds, where $F$ is the frame rate of the camera, the frame rate drops.
In such a case, to still be able to compensate for poor ambient light while ensuring a stable frame rate, the automatic gain controller (AGC) integrated with the measurement unit increases the analog gain used to amplify the measured signal charge.

In contrast to CMOS image sensors, where each individual pixel is equipped with a measurement unit, the signal charge in CCD image sensors is shifted through various components before it is quantized and amplified.
As a result, there are more places where interference can occur.
This means that any noise that occurred before the amplification, such as dark current shot noise, is also amplified.
As a result, a voltage induced by electromagnetic interference will also be amplified, making CCD image sensors more susceptible to interfering signals~\cite{nakamura-2006}.

%% file: 04-ThreatModel/threat_model.tex
\section{Threat Model}

The overarching goal of the adversary is to spoof the image information captured by a CCD image sensor using electromagnetic interference.
Depending on the scenario, the attacker may wish to inject adversarial examples in order to disrupt vision-based intelligent systems for object detection or identification. Alternatively, they may wish to degrade raw images as captured by microscopes or astronomical instruments, to harm research efforts. In a surveillance context, the goal may be to distort images such that further malicious behavior is not recorded accurately. 

We assume that the attacker has knowledge of the target device, sufficient to find technical specifications of the image sensor from a datasheet and access an independent unit to profile for effective signal injection frequencies. 

We assume the adversary has access to off-the-shelf equipment, such as software-defined radios, amplifiers and antennas. The attacker's equipment is assumed to be powerful enough to generate and modulate an attack signal sufficiently quickly to match the performance of the targeted image sensor. 
We also presume that the attacker is capable of generating an arbitrary attack signal.

We assume the attacker can approach the target close enough to mount an attack, for some given transmission power, but since it is an electromagnetic attack, line-of-sight is not required. 

However, under no circumstances can the attacker access the video output of the target camera.
Hence, no synchronization between the attack signal and the camera readout is possible. 
As we discuss later, this condition substantially limits the fidelity with which an attacker can recreate an image at the target; however, we argue that it is by far the most realistic case.

%% file: 05-Attack/attack.tex
\section{Signal Injection Attack}\label{sec:sig_inj_attack}

\begin{figure*}[t]
	\centering
	\includegraphics[width=0.85\textwidth]{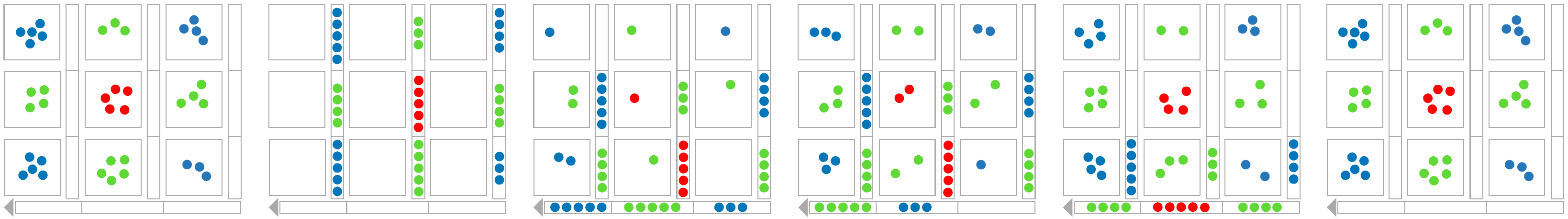}
	\caption{Illustration of a readout of the generated signal charge from an IT-CCD with Bayer color filter array.}
	\label{fig:ccd_readout} 
\end{figure*}
\begin{figure}[t]
	\centering
	\includegraphics[width=0.45\textwidth]{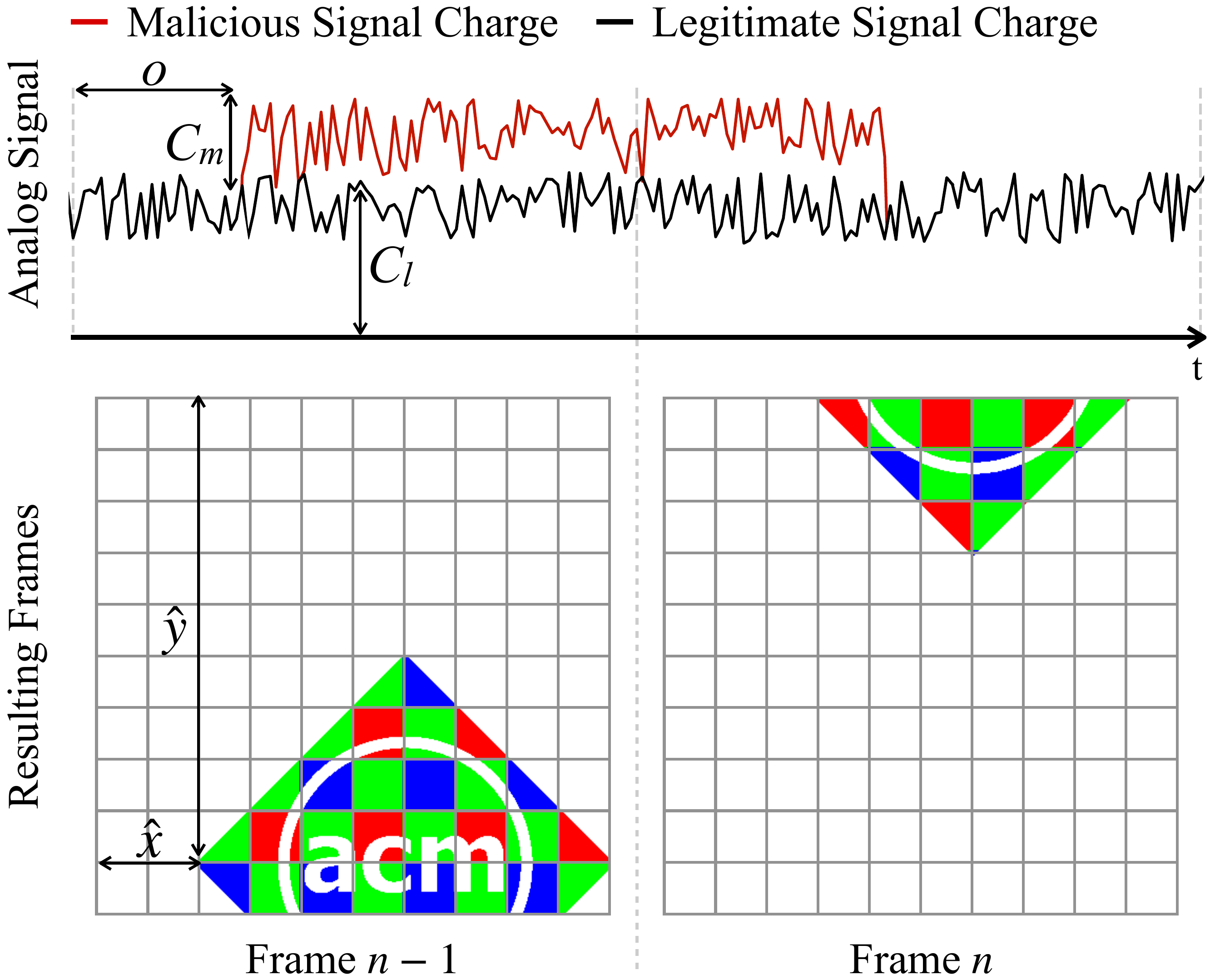}
	\caption{Detailed illustration of how the signal injection attack affects the capturing of frames. The upper part shows the signal charge in the time-domain before it is digitized. After the time offset $o$, the attack signal couples onto the image sensor, increasing the amplitude of the legitimate signal charge $C_l$ by the malicious amplitude $C_m$. The lower part of the figure shows the resulting frames. Due to the misalignment $o$ between the readout and the malicious signal, the induced noise is offset by $\hat{x}$ and $\hat{y}$. As a result, the distortion is stretched along two consecutive frames.}
	\label{fig:ccd_readout_under_attack} 
\end{figure}

Normally, a sensor should only react to the one specific physical stimulus it is intended to capture.
In the case of an image sensor this stimulus is light and the result is the generation of signal charge from photodiodes, which is then measured and digitized. 

It is typical for electronic devices to display some susceptibility to electromagnetic interference, wherein incident electromagnetic radiation induces a voltage in components or connections within the device. In an image sensor, this may lead to the charge that was originally accumulated by the photodiodes subsequently being altered by additional charge due to induced voltages in downstream components. 

The image sensor itself cannot determine whether the signal charge was generated by the photodiode array or resulted from electromagnetic interference that coupled onto the circuit. A malicious actor could leverage these factors and emit electromagnetic waves at the resonant frequency of elements within the target CCD image sensor to induce a voltage and subsequently alter the captured image information.

Our hypothesis is that, due to their architecture, CCD image sensors are \textit{particularly susceptible} to the effects of such post-transducer signal injection attacks. 
Three main architectural factors contribute to this:
\begin{enumerate}
    \item \textbf{long signal charge pathway} -- each of the components through which signal charge is shifted may be affected by incident electromagnetic radiation to facilitate signal injection
    \item \textbf{amplification of signal charge} -- increasing the effect of injected signals prior to amplification
    \item \textbf{serialization of pixels for digitization} -- meaning that injected signals can be targeted to single pixels
\end{enumerate}

Figure~\ref{fig:ccd_readout_under_attack} illustrates the attack being used to inject the ACM logo into an otherwise empty image. The addition of a malicious signal above the legitimate signal can be seen, along with impairments in image reproduction due to lack of synchronization. 

The amount of maliciously induced signal charge depends on the received power of the attack signal. 
Intuitively, a physical signal with higher amplitude induces a greater voltage, which in turn increases the brightness of the resulting frame. The received power is influenced by many factors, such as the attacker's transmission power, the distance, losses during propagation and the efficiency of coupling within the image sensor components. As not all factors are under the attacker's control, they can primarily enhance the effects of the attack either by increasing their transmission power or reducing their distance to the target.

Since the signal charge is amplified and quantized at the last step of the readout process, the maliciously induced signal is also amplified.
Under attack, the total brightness of a pixel, represented by luma $Y$, is the sum of the legitimate signal and the additional induced voltage, and can formally be expressed as:

\begin{equation} \label{eq:total_signal_strength}
 Y = \alpha (C_{l} + C_{m}),
\end{equation}
where $\alpha$ is the amplifier gain set by the image sensor, $C_{l}$ the signal charge captured by the sensing part and $C_{m}$ the maliciously induced signal. It is important to note that the attacker cannot produce a negative value of $C_{m}$ and can thus only increase the brightness of a pixel. Furthermore, the range of the ADC places a cap on the usable values of $Y$, so if the legitimate signal already saturates the brightness, then an induced signal cannot brighten it further. 

In order to gain fine-grained control over the injected noise, the attack signal has to be modulated at a rate equal to the readout rate of the image sensor. 
This means, one symbol of the attack signal corresponds to exactly one sample of the image sensor, or in other words to one pixel.
The attacker can calculate the readout rate for the sensor from datasheet values, or empirical testing, and adjust their transmission rate accordingly. However, it is important to note that while the attacker can match the rate, they cannot synchronize the attack signal to the readout signal in absolute terms --- as, per our threat model, they have no feedback channel for this information. 
This gives rise to a time offset error $o$, between the injected signal and the legitimate signal. 
The offset error manifests variously as a translation within the frame to the offset coordinates $\hat{x}$ and $\hat{y}$, a dispersion of intensity across adjacent pixels and as a color distortion due to color channels being misaligned. We elaborate on this below.

\subsection{Attack Execution}\label{sec:attack_execution}

\begin{figure*}[t]
	\centering
	\includegraphics[width=0.75\linewidth]{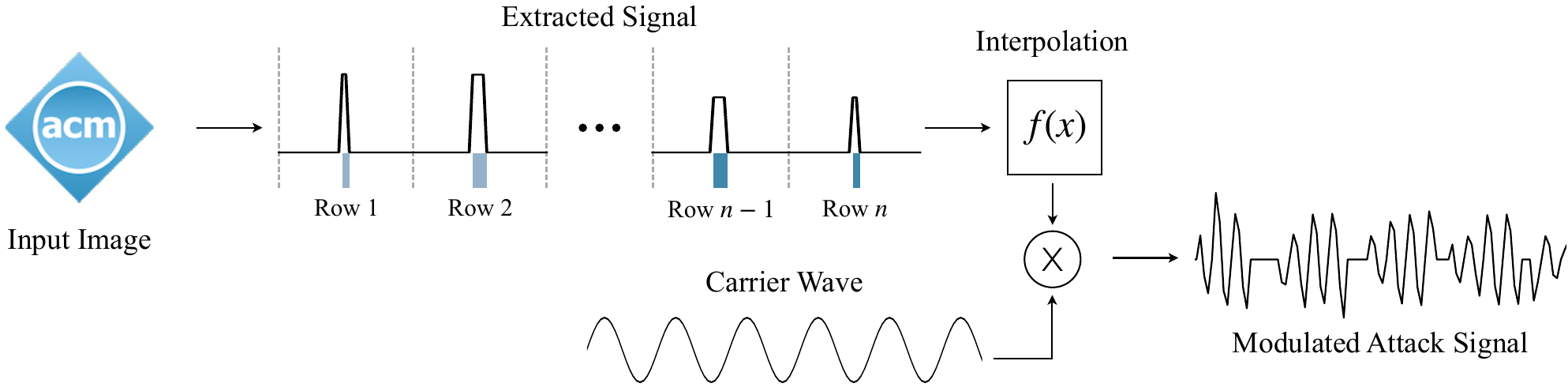}
	\caption{Overview of the necessary steps to generate a malicious attack signal. First, the signal to be transmitted is extracted from the input image by calculating the luminance $Y$ for each pixel. Second, the extracted signal is interpolated to ensure the different sample rates match. Finally, the interpolated signal is modulated onto the carrier wave and transmitted via the software-defined radio. 
	}
	\label{fig:attack_pipeline} 
\end{figure*}

Executing a signal injection attack against a CCD image sensor can be separated into three steps. 
In this section, we will give a detailed overview of the necessary tasks.

In general, any arbitrary data can be modulated onto a carrier wave and induced into the image sensor. In the simplest case, Gaussian white noise can be injected in order to apply random perturbations to the image. However, to demonstrate the possibilities of the attack, we will describe in the following the injection of data in the format of an RGB image, as Figure~\ref{fig:ccd_readout_under_attack} depicts. The content of the injected image can be arbitrary and suited to the scenario, perhaps comprising recognizable patterns or barcodes, masking patterns, or adversarial examples.

\paragraph{Signal Generation}

The origin of the attack signal is a source image of known width and height in RGB format (i.e., three color channels: red, green, blue). In the context of this paper, each pixel of the input image corresponds to one symbol of the attack signal. The pixels are read sequentially from the source image, with an order corresponding to the readout order from the target device. The brightness of the source pixel dictates the amplitude of the modulated attack signal, such that the relative brightnesses are recreated at the target.
In case the resolution of the input image is smaller than the resolution of the target image sensor, padding has to be applied to the attack signal. 
No signal charge is intended to be induced for missing pixels, so the amplitude for these pixels is set to zero.
Likewise, if the input image has a transparent background (alpha channel), the amplitude for transparent pixels is set to zero.

As briefly mentioned in Section~\ref{sec:photodiode_array}, each photodiode only captures the incident light for one wavelength (color).
Assuming that the photodiode at (0,0) only captures light with a wavelength of around 520~nm (green), injecting a malicious signal into the image sensor, while this photodiode is sampled, would intensify the green color channel of the pixel in the final frame.
Since it is not possible to reliably stimulate a specific color channel (owing to the lack of synchronization between the attack and the readout signal), an attacker can avoid the issue by instead seeking to induce different light intensities.
While averaging the intensities of the three color channels can be used to get a rough estimate of the intensity of a pixel, it does not consider the color distribution.
As a result, fine-grained details in the injected image, such as edges, would be lost.
To overcome this issue, we propose to convert the input RGB image into its grayscale version.
This means, the attack signal amplitude is represented by the linear luminance $Y$, which can be calculated using the following equation:

\begin{equation} \label{eq:signal_strength_at}
 Y[x,y] = \frac{0.2126R + 0.7152G + 0.0722B}{255}
\end{equation}
with $x$ and $y$ being the coordinates of the pixel, R, G and, B as the intensities of the respective color channels, the Luma coefficients selected based on the Rec. 709 standard~\cite{bt709}, and a normalization factor to ensure the value remains in the appropriate range.
It should be noted that the attack signal does still stimulate different color channels, and thus the injected distortion will not appear in grayscale.

\paragraph{Resampling/Interpolation}

\begin{figure}[t]
	\centering
	\begin{subfigure}[b]{.499\linewidth}
		\centering
		\includegraphics[width=.95\textwidth]{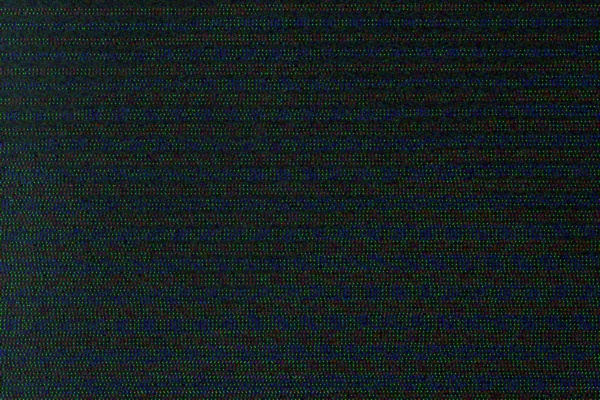}
		\caption{Frame $n-1$}
		\label{fig:frame_1} 
	\end{subfigure}%
	\begin{subfigure}[b]{.499\linewidth}
		\centering
		\includegraphics[width=.95\textwidth]{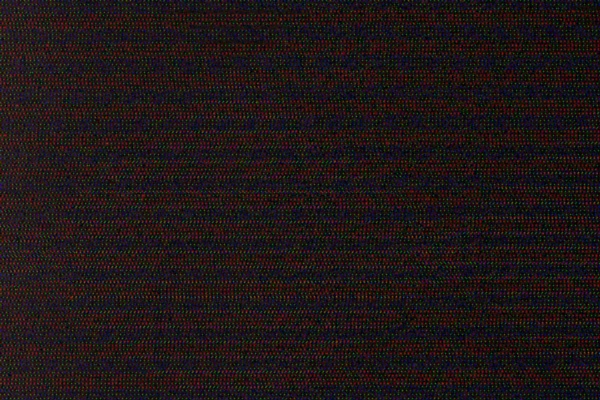}
		\caption{Frame $n$}
		\label{fig:frame_2} 
	\end{subfigure}%
	\caption{Two consecutive example frames captured by the DFM 25G445-ML during the emission of random noise. Due to the wrong sample rate, the injected noise is drifting, causing the stimulus of different photodiodes. As a result, the color of the injected noise changes between consecutive frames.}
	\label{fig:over_under_sampled}
\end{figure}

The modulated symbol rate required to inject arbitrary image information without skipping the readout of some pixels is important. Ideally, the symbol rate of the attack signal should match the readout rate of the image sensor.
Usually, the readout rate is determined by the sample rate of the digital-to-analog converter.
If the symbol rate of the attack signal does not match the readout timings of the image sensor, the injected noise will drift over consecutive frames.

In some cases, the exact sampling rate of the analog-to-digital converter can be obtained from the datasheet of the target camera or image sensor. Sometimes, however, such detailed information is not available and must be calculated instead. The resolution of image sensors is usually specified by two different numbers --- effective and total number of pixels.
Effective pixels represent the number of pixels that are exposed to incident light and used to capture image information. 
In contrast, the number of total pixels specifies the physical size of the image sensor, i.e., the number of photodiodes. 
Usually, the pixels around the edges are light-shielded and used to determine the edges of the captured images, as well as to capture some additional color information about the scene necessary for color calibration.
Since the image sensor reads out all pixels, the number of total pixels has to be used for the calculation.
Based on these facts, the required sample rate can be calculated as follows:

\begin{equation} \label{eq:samp_rate}
 S = N\textsubscript{columns} \cdot N\textsubscript{rows} \cdot F,
\end{equation}
where $N\textsubscript{columns}$ is the width, $N\textsubscript{rows}$ the height and $F$ the frame rate of the image sensor.

For software-defined radio transmitters, arbitrary sample rates may not be possible and hence the attack signal must be appropriately resampled to match the transmission rate, using standard interpolation methods. 

It should be noted that with increasing frame rate and resolution, the required sample rate of the software-defined radio is increasing too.
Depending on the target camera, this may increase the difficulty for the attacker to inject fine-grained distortions.
Figure~\ref{fig:over_under_sampled} shows the results of under- and oversampling, as a direct reproduction of the output image from the camera\footnote{For clarity in print, artificially brightened versions are also given in Appendix~\ref{apx:additional_figures}}.

\paragraph{Transmission}

Once the signal has been extracted and interpolated, it can be transmitted.
Given that the amplitude of the attack signal determines the amount of electric charge induced into the image sensor, the input image is modulated onto the carrier wave using amplitude modulation.
An end-to-end representation of the three attack steps is depicted in Figure~\ref{fig:attack_pipeline}.

%% file: 06-Evaluation/evaluation.tex
\section{Evaluation}

We evaluated the susceptibility to intentional electromagnetic interference of two different CCD image sensors.
In this section, we describe our method and present the results. 

\subsection{Experimental Setup}\label{sec:experimental_setup}

\begin{figure}[t]
	\centering
	\includegraphics[width=0.95\linewidth]{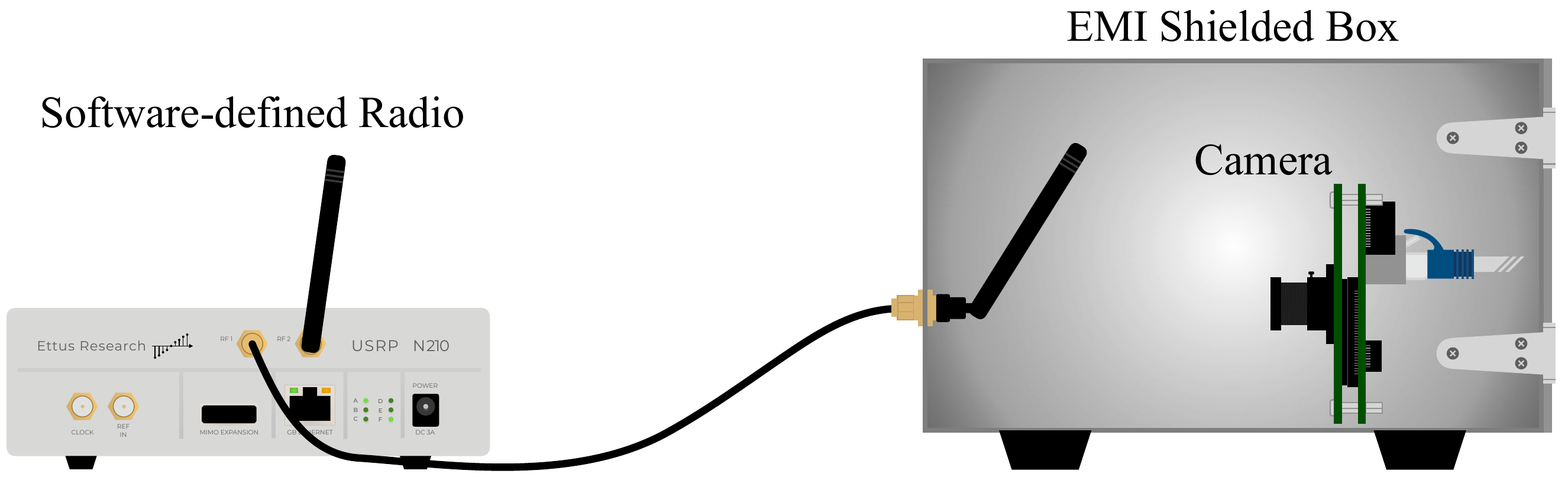}
	\caption{Experimental setup used for our evaluation. The camera was placed inside an EMI shielded box to prevent interference with and from other devices.}
	\label{fig:experimental_setup} 
\end{figure}

We arranged equipment to examine the effects of interference signals on CCD cameras. An overview of the experimental setup is given in Figure~\ref{fig:experimental_setup}. The camera under test was placed inside an RF shielded box, along with the transmission antenna, at a fixed distance $d$. The camera was connected to a desktop PC, to capture the image output. The attack transmitter was an Ettus Research USRP N210 software-defined radio, driven over Ethernet by the same desktop PC, running Ubuntu 18.04 and GNURadio 3.8. The USRP was equipped with an UBX-40 daughterboard, which provides a maximum output power of 100~mW~\cite{ubx_daughterboard}. The antenna was an omnidirectional monopole exhibiting 3 dBi gain and optimised for transmission at 900 MHz.

Two different CCD cameras were tested, namely a DFM 25G445-ML and a 420TVL CCTV board camera~\cite{amazon-sony-tvl}. The DFM 25G445-ML is a professional GigE color board camera used in a wide variety of applications, for instance industrial automation, quality assurance and surveillance~\cite{dfm25g445}, and is equipped with a Sony ICX445AQA image sensor~\cite{icx445aqa}. 
This camera was directly connected to the desktop PC via a shielded Ethernet cable (S/FTP).
In contrast, the CCTV board camera uses an unspecified 1/3\,'' CCD Sony image sensor and only provides an analog composite video output.
Such analog image sensors can often be found in older CCTV cameras or cheap drones. The analog image output was passed through a VHS-to-USB capture device~\cite{vhs_to_usb} to digitize the video signal before it was delivered to the PC. 

The use of an RF shielded box ensured that the experiments were not corrupted, either by outside signal sources, or by the attack signal affecting components downstream of the CCD camera itself. Its presence also ensured that we were compliant with relevant regulations on use of radio spectrum. 
To validate that the attack signal was not induced into downstream components, such as cabling or the VHS-to-USB capture device, we tested the attack with the camera switched off, in which case no effects were observed. 

The RF shielded box also provided a controlled lighting environment, providing a dark scene that was not affected by variations in ambient light outside. This allowed a more accurate measurement of the impact of the attack. 
Under normal operation all the captured video frames were almost entirely black.
Only some pixels were colored due to the various types of noise, such as readout and dark current shot noise, generated by the camera itself~\cite{irie2008model, widenhorn2002temperature}. 

The shielded box provides only an RS232 port for cable pass-through, so in order to connect the cameras to the PC, we routed the connections via shielded RS232 adapters. For the DFM 25G445-ML we used an RJ45 to RS232 adapter, while the analog CCTV camera was connected via an RS232 breakout adapter.

To measure the impact of any given attack signal, we captured a series of video frames with the attack signal off (``legitimate frames'') or on (``malicious frames''). 
We collected three legitimate frames and seven malicious frames. 
The frames were then compared using the Structural Similarity Index Measure (SSIM)~\cite{wang2004image}. 
The SSIM compares two images and results in a high value if they are similar or a low value if they are dissimilar\footnote{The source code for our evaluation is available at \url{https://github.com/ssloxford/ccd-signal-injection-attacks}}. 
As each camera under test was observing a controlled scene, any dissimilarity in the output image can be considered the result of the injected attack signal (along with a small amount of random sensing noise). Indeed, for consecutive legitimate frames, the SSIM remained consistently high and close to a value of 1.0 (only dipping below due to sensing noise). Each malicious frame was compared with each legitimate frame to produce 21 SSIM values and the mean value taken. This averaging not only reduced the effect of sensing noise, but also that of injected distortions affecting each frame differently due to a lack of synchronization. Along with the SSIM, other image quality metrics were collected and are presented in Appendix~\ref{apx:additional_results}. We focus on the SSIM throughout this evaluation, noting that the results of each metric were similar and would only influence parameter values rather than any procedural change. 

\subsection{Carrier Frequency $f_c$}\label{sec:carrier_freq}

Components within the image sensor will be most susceptible to signal injection near their resonant frequencies. 
While it is theoretically possible to model and calculate the likely resonant frequencies for a target image sensor, many factors influence the calculation, making it non-trivial and error-prone. 
In this section, we describe empirical testing used to determine the most effective carrier frequency for the two tested cameras.

\subsubsection{Method}

We captured video frames while performing a frequency sweep with the transmitter; ranging the carrier frequency from 50 to 5000~MHz in a step size of 1~MHz. At each step, the transmitter modulated a 1~kHz sine wave onto the carrier at a sample rate of 25~MSPS. For this experiment, the antenna distance $d$ was set at approximately 3~cm and the output power of the software-defined radio was set to the maximum (20.1~dBm, $\sim$100~mW).

For both cameras, the settings were set to \textit{auto}, which means the exposure time and gain were automatically set by the camera itself.

Ten frames were captured (three legitimate, seven malicious) and compared using the SSIM metric, as described above. The most effective carrier frequency was selected based on the smallest SSIM value.
In other words, the frequency that caused the smallest SSIM values must have induced the most significant perturbations. 
On the other hand, an ineffective carrier frequency did not induce any signal charge and led to high SSIM values similar to those measured between legitimate frames.

\subsubsection{Results}

\begin{figure*}[t]
	\centering
	\begin{subfigure}[b]{.499\linewidth}
		\centering
		\includegraphics[width=.9\textwidth]{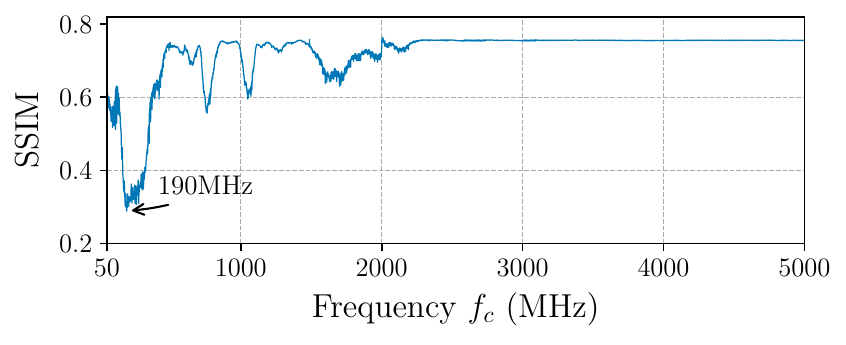}
		\caption{DFM 25G445-ML}
		\label{fig:25G445} 
	\end{subfigure}%
	\begin{subfigure}[b]{.499\linewidth}
		\centering
		\includegraphics[width=.9\textwidth]{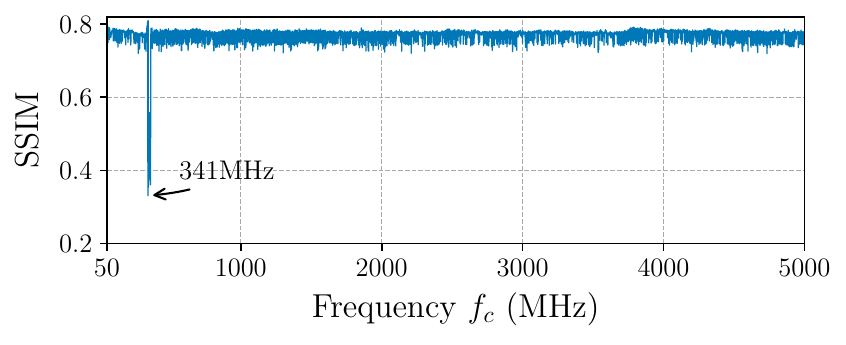}
		\caption{Analog CCTV board camera}
		\label{fig:vhs} 
	\end{subfigure}%
	\caption{Results of the frequency sweep. The SSIM represents the similarity between the frames captured during normal operation and while an attack signal at the carrier frequency $f\textsubscript{c}$ was emitted.}
	\label{fig:ssim_results}
\end{figure*}

The results of the frequency sweep for both cameras, the DFM 25G445-ML and the analog CCTV board camera, are visualized in Figure~\ref{fig:ssim_results}.
As the graphs show, the most effective carrier frequency was 190~MHz for the DFM 25G445-ML and 341~MHz for the analog CCTV board camera. At these respective frequencies the distortion level was comparable for each camera, with SSIM values below 0.4 in both cases. However, the range of effective frequencies was different in each case. For the DFM 25G445-ML, a wide range of frequencies had a noticeable effect on the image, while the analog camera exhibited only a small range of effective frequencies. For both cameras, the highest SSIM values were already quite low, below 0.8 consistently. This is due to the cameras being set in \textit{auto} mode, causing them to increase exposure and gain settings in an attempt to compensate for the dark environment in the shielded box. The impact of sensing noise is increased under these circumstances. 

These results indicate that an attacker could inject a malicious signal for either camera and affect the output image substantially. The freedom for an attacker to select a convenient transmission frequency depends on the target camera. 

Due to space constraints and the option to precisely control the camera parameters, such as exposure and gain, which allows us to evaluate the attack under controlled conditions, the rest of the paper will focus on the evaluation and results of the DFM 25G445-ML. Nevertheless, since we know that the analog CCTV board camera is also vulnerable to signal injection attacks under the same attack settings, the following findings will be applicable to it too.

\subsection{Transmission Power}

\begin{figure}[t]
	\centering
	\subcaptionbox*{} {
		\includegraphics[width=1\linewidth]{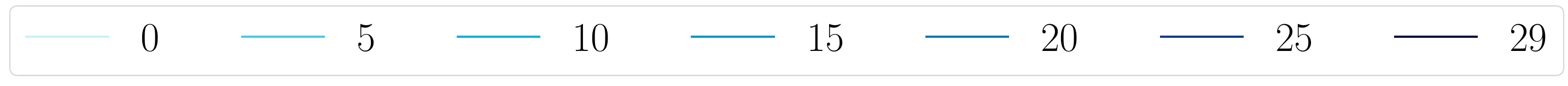}\vspace{-4mm}%
	}
	\subcaptionbox{SSIM vs. Transmission Power $P_{out}$\label{fig:ssim_vs_pout}}{%
      \includegraphics[width=0.45\textwidth]{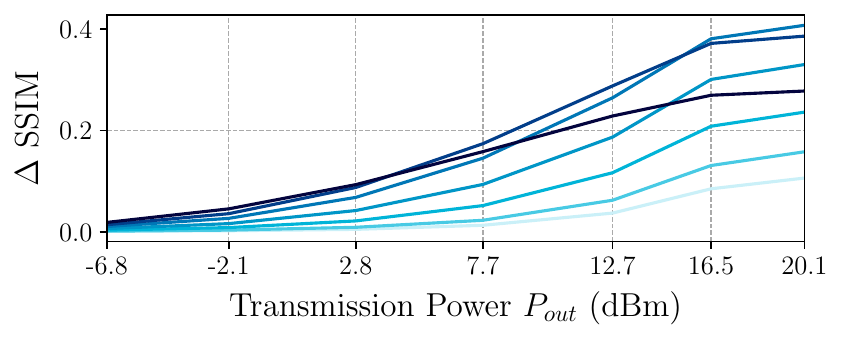}%
      }\par\medskip
    \subcaptionbox{SSIM vs. Distance\label{fig:ssim_vs_distance}}{%
      \includegraphics[width=0.45\textwidth]{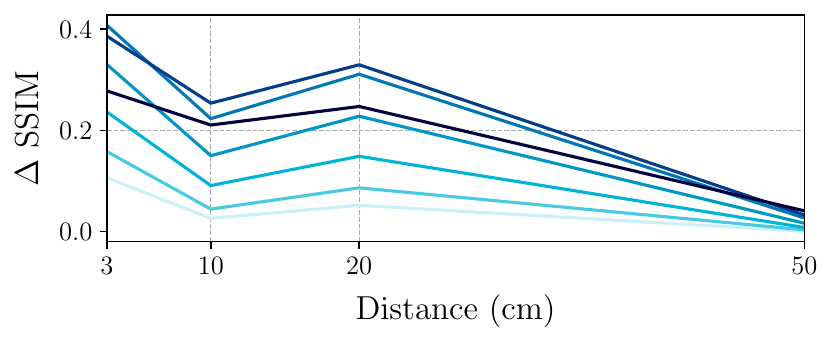}%
      }      
	
	\caption{Evaluation results for different camera gain settings. The upper part depicts the relationship between the transmission power $P_{out}$ and the amount of noise induced into the DFM 25G445-ML from a fixed distance of 3~cm. The lower part illustrates how the injected noise diminishes with increasing distance between the camera and the transmitter for the maximum transmission power of 20.1~dBm.}
	\label{fig:ssim_results_power_distance}
\end{figure}

The transmission power of the malicious signal is a decisive factor for the success of the attack.
Depending on the environment, target camera and its settings, the minimum required power varies.
In this section, we present an analysis of the relationship between signal strength and the amount of induced distortions for the DFM 25G445-ML.

\subsubsection{Method}

To determine the minimum required transmission power under different settings and to evaluate the relationship between the required output power and the amount of perturbations, we tested the DFM 25G445-ML at a fixed distance of 3~cm from the transmitting antenna. 
In accordance with the results of the frequency sweep, we set the carrier frequency to 190~MHz and then varied the transmission power from -6.8~dBm to 20.1~dBm (output power at $f_c = 190$~MHz as measured with an oscilloscope).
We repeated the experiment for a range of camera gain settings. Under real-world conditions, the ambient lighting of the environment would influence the exposure and gain settings of the camera. However, there is an upper limit on the exposure value that can be used without reducing the frame rate. In low light level environments, for example indoors with only artificial lighting, it is highly likely that a camera will increase the gain of the amplifier in the measurement unit to compensate for the low ambient brightness. For all experiments the exposure time was set to the smallest possible value (10~$\mu$s), but the gain values ranged from 0 to 29 (unitless values as offered by the camera control software).

As with the previous experiments, we collected ten frames, three legitimate and seven malicious, and calculated the SSIM between them. However, as the gain increases, the noise floor in the captured frames increases, resulting in a lower structural similarity, even between legitimate frames.
To circumvent this issue and to facilitate the comparison of the results, we calculated the change in SSIM values, $\Delta$ SSIM:

\begin{equation} \label{eq:delta_ssim}
 \Delta SSIM = SSIM_{legitimate} - SSIM_{malicious}.
\end{equation}
where $SSIM_{legitimate}$ is calculated among legitimate frames only and $SSIM_{malicious}$ is measured between legitimate and malicious frames.
This approach makes it easier to identify the additional interference caused by the signal injection attack.

\subsubsection{Results}

As can be seen in Figure~\ref{fig:ssim_vs_pout}, for a high image sensor gain, the signal strength of only -2.1~dBm was sufficient to induce a malicious signal charge into the CCD image sensor of the DFM 25G445-ML.
The results indicate that under such advantageous conditions for the adversary, increasing the transmission power to 20.1~dBm ($\sim$100~mW) can induce considerable noise.

Remarkably, however, the output power of the USRP was even enough to cause distortions when the amplifier of the measurement unit was switched off. This suggests that the attacker is not reliant upon vulnerable configuration within the target camera, as long as they are able to increase their transmission power to compensate. 
Indeed, increasing transmission power, rather than relying on high image sensor gain, is beneficial for the attacker. 
We observed, somewhat surprisingly, that with the highest image sensor gain of 29, the injected noise level was lower than with a gain of 25. This was because at such a high gain, the low level sensing noise was amplified such that it already saturated some pixels --- making it impossible to further increase the signal charge for these pixels via the signal injection attack. Instead, if the gain is kept low, the attacker inject their signal with a high signal-to-noise ratio.

\subsection{Attack Distance}

While the attack signal propagates through space, it is attenuated.
Therefore, the effect of the attack diminishes with increasing distance between the target camera and the malicious transmitter.
In the following, we evaluated the feasibility of the attack for different distance settings. 

\subsubsection{Method}

We used the same experimental setup as described previously and depicted in Figure~\ref{fig:experimental_setup}.
However, this time we fixed the transmission power of the USRP to the maximum (20.1~dBm, $\sim$100~mW) and only varied the distance between the camera and the transmitting antenna.
We emitted a sine wave with a frequency of 1~kHz that was modulated onto a carrier wave with $f_c = 190$~MHz.
In line with previous experiments, we collected three legitimate and seven malicious frames for each distance setting.
Due to the size limitations of the shielded box, we had to restrict the evaluated distances to 3, 10, 20, and 50~cm.
Again we calculated $\Delta$ SSIM between legitimate and malicious frames.

\subsubsection{Results}

Consistent with our expectations, Figure~\ref{fig:ssim_vs_distance} shows that, as the distance between the target and the transmitter increased, the amount of induced signal charge decreased.
The results indicate that the transmission power of the USRP without an amplifier is not sufficient to inject distortions into the frames from more than 50~cm away.
In such a setting, not even a high camera gain, the most beneficial setting for the adversary, is enough to cause substantial image distortions. While this represents the limits of our experimental setup, higher received power could be stimulated straightforwardly by an attacker, in order to improve the range of the attack --- either by increasing transmission power or employing a directional antenna with higher gain.  
Based on the same assumptions as in \cite{kune2013ghost}, we can utilize the Friis transmission equation to roughly estimate the requirements for an attack at a certain distance $d$:

\begin{equation} \label{eq:friis}
 P_r = P_t + G_t + G_r + 20\log_{10}(\frac{\lambda}{4\pi d}),
\end{equation}
where $P_t$ is the transmission power of the attack signal in dBm, $G_t$ and $G_r$ are the antenna gains of the transmitting and receiving antennas in dBi and $\lambda$ is the wavelength corresponding to the selected carrier frequency $f_c$. 

Interestingly, the amount of induced noise from 20~cm away exceeded the amount coupled onto the circuit from a distance of only 10~cm.
This result can be explained by the fact that the antenna was aligned at a different angle to the camera, i.e., the camera was placed slightly outside the radiation pattern of the antenna.

\subsection{Evaluation of CMOS Image Sensors}

Our hypothesis is that CCD image sensors are vulnerable to intentional EMI due to their architecture.
As elaborated in Section~\ref{sec:background}, the single measurement unit makes CCD image sensors more susceptible to noise.
In contrast, CMOS image sensors that have a measurement unit in each pixel would not be expected to be as susceptible, since the EMI has less opportunity to couple onto the image sensor before the amplification process. 
We conducted an experiment to verify this expectation. 

\subsubsection{Method}

We repeated the carrier frequency experiment described in Section~\ref{sec:carrier_freq}, using two CMOS cameras instead of CCD units. 
We tested a Logitech C922, a widely used webcam, and an Axis M3045-V semi-professional dome surveillance camera.
As with the CCD cameras, we specifically chose these cameras based on their capability to manually adjust camera settings such as gain and exposure. For each camera we selected an attack distance of 3~cm and performed a carrier frequency sweep from 50 to 5000~MHz with a step size of 1~MHz, while modulated by a 1~kHz sine wave. We again captured three legitimate and seven malicious frames and computed the SSIM values. 

\subsubsection{Results}\label{sec:cmos_results}

\begin{figure*}[t]
	\centering
	\begin{subfigure}[b]{.499\linewidth}
		\centering
		\includegraphics[width=.9\textwidth]{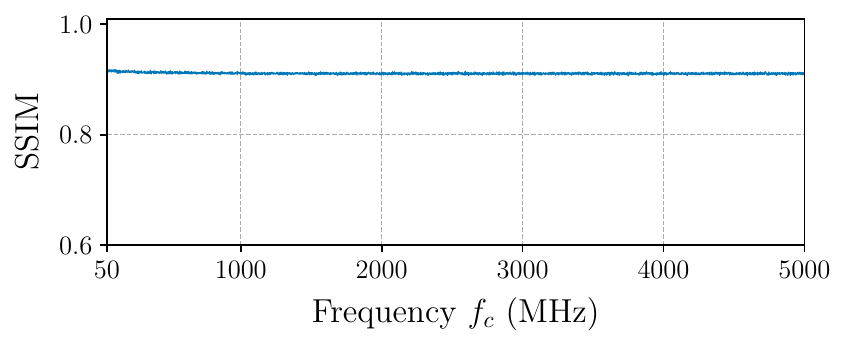}
		\caption{Axis M3045-V}
		\label{fig:axis_mv3045} 
	\end{subfigure}%
	\begin{subfigure}[b]{.499\linewidth}
		\centering
		\includegraphics[width=.9\textwidth]{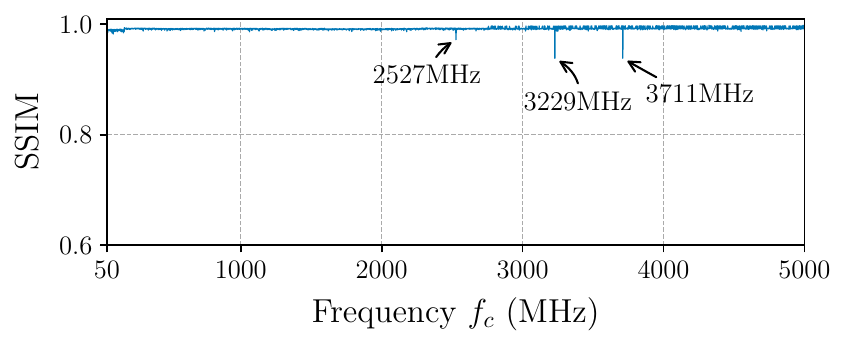}
		\caption{Logitech C922}
		\label{fig:logitech_c992} 
	\end{subfigure}%
	\caption{Results of the frequency sweep for the two tested cameras with CMOS image sensors. The SSIM represents the similarity between the frames captured during normal operation and while an attack signal at the carrier frequency $f\textsubscript{c}$ was emitted.}
	\label{fig:cmos_freq_sweep}
\end{figure*}

In Figure~\ref{fig:cmos_freq_sweep}, the results of the frequency sweep for the two tested cameras are presented.
As expected, the SSIM values for both cameras are consistently high, except for a few occasions for the Logitech C922.
We inspected and tested the carrier frequencies for the unexpectedly low SSIM values manually to investigate if signal injection attacks at these frequencies would be possible.
However, all attempts were unsuccessful.
We suspect that the reason for these outliers was most likely a problem in the communication between the camera and the PC it was connected to.
The results indicate that the tested CMOS image sensors are not susceptible to electromagnetic emanation.
At the same time, they also confirm our hypothesis that signal injection attacks against CCD image sensors are uniquely a product of their architecture. This does not necessarily rule out other EMI attacks on CMOS image sensors under different circumstances, but we consider that to be beyond the scope of this paper. 

\subsection{Fine-Grained Control}

\begin{figure}[t]
	\centering
	\includegraphics[width=1.0\linewidth]{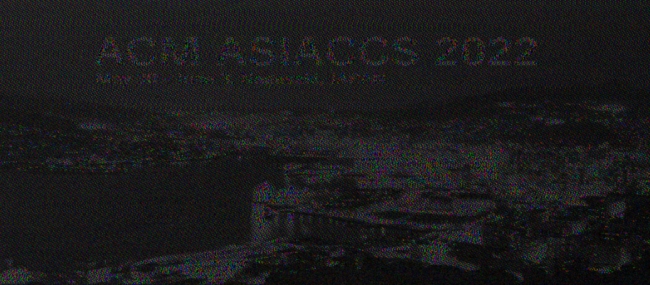}
	\caption{Example of a signal injection attack, illustrating the fine-grained control an adversary can gain over the captured frame.}
	\label{fig:signal_injection_example} 
\end{figure}

In this section, we show how an adversary can exploit fine-grained control over the captured image information. 

\subsubsection{Method}

We replicated the experimental setup as previously described and depicted in Figure~\ref{fig:experimental_setup}.
The camera was placed 3~cm away from the transmitting antenna, and the image sensor gain was set to 29.
We then executed the attack by following the steps described in Section~\ref{sec:sig_inj_attack}. In the case of the DFM 25G445-ML, it was easy to infer from the datasheet that the sample rate of the underlying Sony ICX445AQA image sensor is 36~MHz. Based on the earlier results, the transmission was made with ${f_c = 190}$~MHz and peak transmission power of 20.1~dBm.

\subsubsection{Result}

The resulting frame is shown in Figure~\ref{fig:signal_injection_example}. 
The banner image for the AsiaCCS 2022 conference is visible, with both the text and cityscape identifiable to the human eye\footnote{For clarity in print, we give an artificially brightened version of the image in Appendix~\ref{apx:additional_figures}.}.
To demonstrate the possibilities of such fine-grained injections, we uploaded the frame with the injected logo to the Google Cloud Vision API~\cite{google_vision_api}, which correctly recognized the text ``ACM ASIACCS 2022''.
The output from the API request is depicted in Figure~\ref{fig:google_api_result} in the Appendix.

\subsection{Use Case: Barcode Scanning}

To illustrate an end-to-end attack, we consider the case of automated barcode scanning in manufacturing or logistics. 
One major advantage of CCD image sensors is the global shutter behavior, meaning that all pixels are exposed and read out at the same time, allowing the capture of images that are free from geometric distortion, even when moving quickly. For this benefit, in combination with their high sensitivity, CCD image sensors can often be found in cameras used for barcode scanning in warehouses~\cite{sick-autoident-2014,weng2012design}. This process plays a crucial role in tracking items through industrial processes and accounting for inventory. We consider an attack that seeks to remotely disrupt the performance of the barcode scanning, thereby either inhibiting the efficient flow of tracked items or corrupting the inventory management of the facility. As automated CCD barcode scanners often handle hundreds of barcodes per second~\cite{allied-autobarcode-2021}, even a short attack can quickly impact a large number of items. 

Scanning a barcode relies on the color contrast between bright and dark bars.
In this section, we show how a malicious signal can break this contrast and cause the barcode reading to fail.

\paragraph{Experimental Setup}

Similar to the experimental setup in the previous section, we placed the DFM 25G445-ML in the RF shielded box 3~cm away from the transmitting antenna.
However, additionally, we placed a cardboard box with two barcodes together with a light inside the box.
The camera parameters were set to match the settings that would be chosen by the automatic exposure and automatic gain control for an indoor environment with artificial ceiling lighting, as it can be found, for instance, in a warehouse.
To be more precise, we tested exposure times ranging from 20,000~$\mu$s to 33,000~$\mu$s and image sensor gains from 0.0 to 8.7.
The camera was connected to a PC running a Python script that captured frames with a frame rate of 30~FPS.
The captured frames were analyzed for barcodes using the popular library \textit{pyzbar}.
If a barcode was detected, we stored the decoded data in a CSV file.
For each parameter configuration we collected 1,000 frames --- 500 under normal operation and 500 while emitting random noise at ${f_c = 190~MHz}$.
The transmission power of the USRP was again set to the maximum (20.1~dBm).

\paragraph{Results}

\begin{figure}[t]
	\centering
	\begin{subfigure}[b]{.499\linewidth}
		\centering
		\includegraphics[width=.95\textwidth]{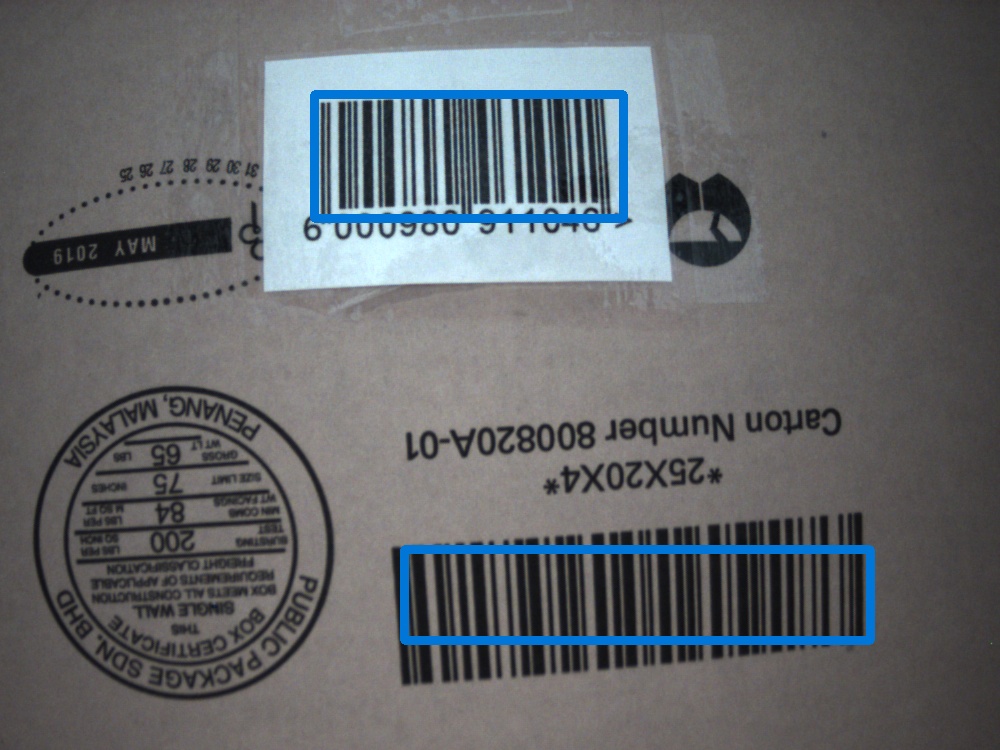}
		\caption{Normal operation}
		\label{fig:barcode_legitimate} 
	\end{subfigure}%
	\begin{subfigure}[b]{.499\linewidth}
		\centering
		\includegraphics[width=.95\textwidth]{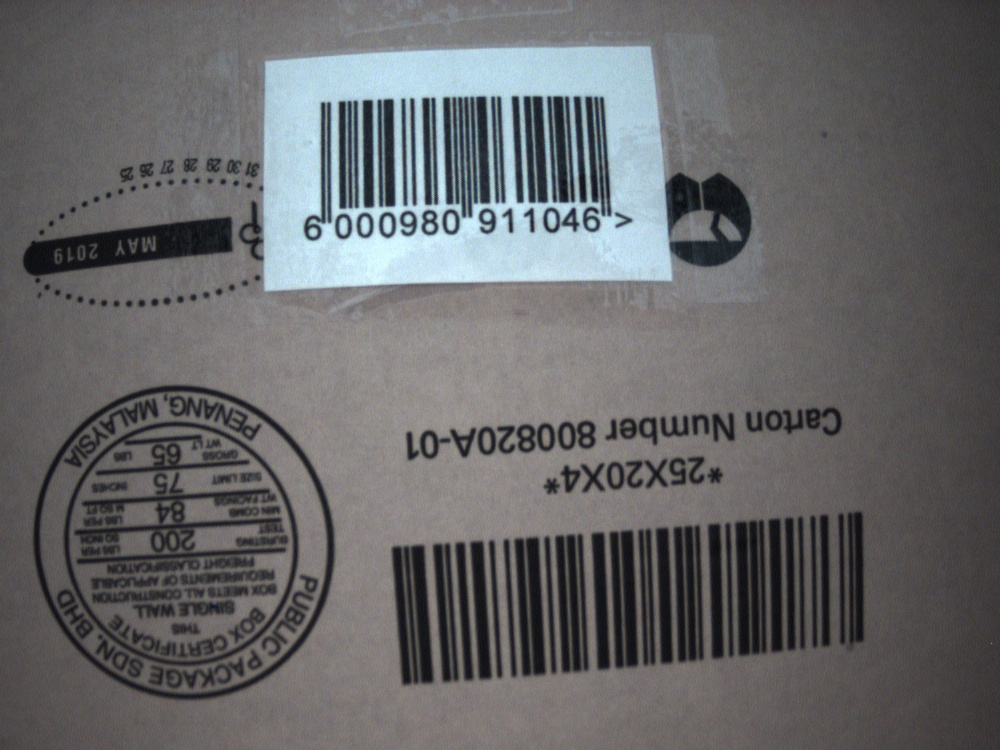}
		\caption{Under attack}
		\label{fig:barcode_malicious} 
	\end{subfigure}%
	\caption{Example frames from the barcode scanner captured by the DFM 25G445-ML. The left frame was captured during normal operation, while the right frame was collected during the transmission of a malicious signal. The injected noise is barely noticeable, but causes the detection of the barcodes to fail.}
	\label{fig:barcode_results}
\end{figure}

\begin{figure}[t]
	\centering
	\includegraphics[width=1.0\linewidth]{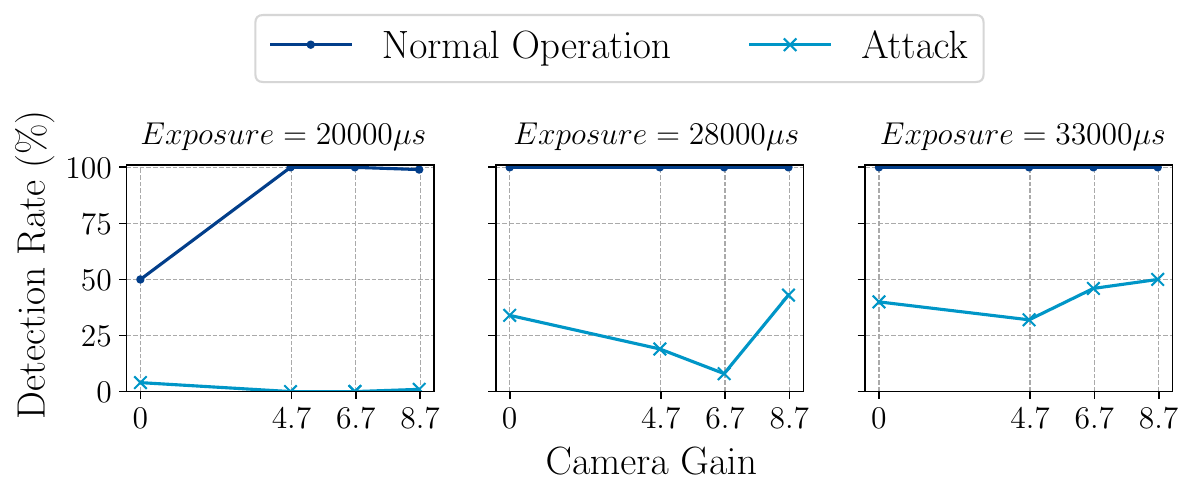}
	\caption{Results of the automatic barcode scanning under different lighting settings.}
	\label{fig:barcode_vs_exp} 
\end{figure}

The results of our experiments clearly show that injecting random noise into a CCD image sensor used for barcode scanning can substantially reduce the reliability of the scanning system.
In Figure~\ref{fig:barcode_vs_exp}, we present the results for different camera settings.
For the lowest selected exposure of 20,000~$\mu$s and no additional amplification of the signal charge the captured frames were slightly underexposed.
This led even during normal operation to a detection rate of only 50\%. 
As such, it is not surprising that the injected noise reduced the detection rate even further.
However, increasing the exposure time and the gain improved the performance under normal operation significantly, leading to a consistent detection rate above 99\%.
At the same time, the attack effectiveness diminished with increasing exposure time, and contrary to our expectations, for higher gains.
This observation can be explained by the increasing contrast between the white background and the black bars of the barcode.
Nevertheless, under optimal settings, for instance, for ${Exposure = 20000~\mu}$s and a gain of 8.7, the attack caused the detection rate to drop to 1\%.
In Figure~\ref{fig:barcode_results}, two example frames, captured during normal operation and while emitting the attack signal, are depicted.

%% file: 07-Limitations/limitations.tex
\section{Limitations}

In this section, we provide an overview of the limitations of our evaluation and the signal injection attack itself.

\subsection{Evaluation Limitations}

In this paper, we presented the first approach of a post-transducer signal injection attack against CCD image sensor to raise the awareness about such attacks and their potential impact.
However, our evaluation has some limitations.

First, we evaluated the signal injection attack on two different commercial off-the-shelf cameras equipped with CCD image sensors.
However, as mentioned earlier, CCD image sensors are often used in scientific, professional, or even military applications.
Although we believe that the results of this paper are applicable to any other CCD image sensor, it is not guaranteed that image sensors used in such specialized applications will not be better protected.

Second, as described in Section~\ref{sec:experimental_setup}, we only had access to a small RF shielded enclosure.
The dimensions of the box constrained the setup of our experiments to short distances between the malicious transmitter and the target camera.
As a result, we only investigated the attack success for short distances up to 50~cm. 
In addition, we only tested the attack with a low output power of 100~mW maximum to ensure that we comply with local regulations and our signal does not interfere with any legitimate communication channels.

\subsection{Attack Limitations}

The signal injection attack in its most basic approach, as it is presented in this paper, has some limitations that are difficult to work around and have to be taken into account by the adversary.

Our threat model assumes that the adversary cannot access the video feed of the target camera.
As described in Section~\ref{sec:sig_inj_attack}, it is therefore not possible to synchronize the readout of the signal charge with the attack signal.
This lack of synchronization introduces two major issues.
First, the injected perturbation appears at random locations ($\hat{x}$,$\hat{y}$), making it impossible for the adversary to target specific parts of the frame.
Second, as we showed in Section~\ref{sec:attack_execution}, it is not possible to stimulate a certain color channel.
Nevertheless, depending on the intentions of the adversary, these limitations might not be important.
For instance, if the goal is to fool an object detection algorithm, the attacker could draft adversarial examples that are effective independent of the injected location~\cite{sayles2020invisible}. 
We consider our assumptions realistic but limiting and highlight that if the adversary could obtain synchronization, by monitoring the camera output, they would enjoy far greater control over the final image.

%% file: 08-Countermeasures/countermeasures.tex
\section{Countermeasures}

Countermeasures to protect sensors from signal injection attacks can be divided into two categories --- attack prevention and attack detection.
In this section, we discuss various approaches for both categories in the context of intentional EMI against CCD image sensors.

\subsection{Attack Prevention}

The prevention of intentional EMI is challenging and can often be seen as an arms race between the defender and the attacker.

\subsubsection{Shielding}

The most obvious solution to prevent a malicious signal from coupling onto the image sensor circuit is shielding.
However, sensors that have to interact with their surroundings are not easy to shield.
For example, in the case of an image sensor, light has to reach the photodiode array.
While it is possible to add a fine metallic mesh in front of the sensing part, it diminishes the quality of the captured frames and only provides limited protection.
Furthermore, shielding affects the airflow and the thermal dissipation.
This is especially disadvantageous for CCD image sensors since higher temperatures cause the generation of more dark current~\cite{widenhorn2002temperature}. 
Moreover, retrofitting the camera with additional shielding is expensive, time-consuming, and potentially not even possible.
Finally, shielding cannot fully protect from malicious electromagnetic waves.
Although shielding does attenuate the induced signal, the effectiveness depends on the thickness of the shield~\cite{tong2016advanced}.
A sophisticated adversary with powerful equipment might still be able to emit a signal that can penetrate the shielding and couple onto the target image sensor.

\subsubsection{Camera Redundancy}

Another straightforward protection approach is the usage of multiple cameras. 
Ideally, the second camera is equipped with a different image sensor model to reduce the likelihood that its circuitry will respond to the same resonant frequency.
Nevertheless, adding camera redundancy significantly increases the costs and provides only limited improvement in protection, as the attacker can still target both image sensors.

\subsection{Attack Detection}

Recent academic research has proposed multiple approaches to detect signal injection attacks against different types of sensors.
In comparison to attack prevention mechanisms, detection approaches can often be implemented in software and are easier to deploy retrospectively.

\subsubsection{Dummy Sensor}

Similar to camera redundancy, the authors in~\cite{tu2021transductionasiaccs} proposed a kind of sensor redundancy.
However, the second sensor, which should be placed directly next to the sensor to be protected, is only a so-called dummy sensor.
It is a duplicate of the original sensor with exactly the same circuit and properties to ensure that it responds to the same resonant frequency.
However, it does not have a sensing part.
Therefore, no voltage can be generated through an external, physical stimulus leading to a sensor output of always 0~V.
In case the microcontroller can measure a voltage at the sensor output, the signal was potentially injected via a post-transducer signal injection attack.
Since the observed signal from the dummy sensor is the raw attack signal, it can be used to correct the signal measured by the original sensor.

Although this approach could be applied to CCD image sensors, it has various drawbacks.
The major disadvantage is that the total size of the image sensor would become twice as large.
 This is in particular an issue for larger sensor arrays composed of multiple CCD image sensors, as they are used in telescopes and satellites, since doubling the size of the sensor array would be impractical.
Moreover, such image sensors are highly complex with sophisticated architectures making them very expensive.
Producing a dummy sensor with the exact same properties will almost certainly not only double the size, but also the price.
Finally, it is not guaranteed that the resonant frequency of the dummy sensor matches the one of the original sensor.

\subsubsection{Modulating the Sensor}

The detection mechanism PyCRA proposed by~\cite{shoukry2015pycra} is another promising approach to detect signal injection attacks.
The idea is similar to the previously described dummy sensor.
If an active sensor, i.e., a sensor with an emitter and a sensing part, for example, a Light Detection and Ranging (LiDAR) sensor, does not emit a signal, the sensing part should not be able to measure a response.
If this is still the case, the probability that the signal is not authentic is relatively high and an alarm can be raised.
The disadvantage of this method is that it is tied to active sensors.
To circumvent this limitation,~\cite{zhang2020detection} introduced the idea of sensor modulation.
A passive sensor only senses a physical property and outputs a voltage when it is powered.
If the sensor is switched off, no voltage should be present on the sensor output and the microcontroller should measure 0~V.
For both of the discussed detection mechanisms, it is impossible for an attacker to inject a malicious signal without being detected if the sensors or the emitter are turned on and off in an unpredictable, random sequence.

\subsubsection{Adapting existing Detection Mechanisms}

For the detection of post-transducer signal injection attacks against CCD image sensors, we can adapt the two aforementioned detection techniques.
Since image sensors do not have an emitter, we can only control the sensing part.
However, the photodiode array is always on, which means a signal charge is generated as soon as light falls onto it.
Therefore, instead of turning the sensor on and off, we propose reducing the exposure time of the image sensor to the lowest possible value for the duration of a single frame in an unpredictable, random sequence.
Due to the very short exposure time, none or minimal signal charge should be generated.
As in \cite{shoukry2015pycra} and \cite{zhang2020detection}, if the sensor still captures a signal, then it is highly likely that the voltage was caused by EMI coupling onto the circuit.
The main advantage of this method is that it can be implemented in software and deployed retrospectively.
Unfortunately, this approach also has drawbacks.
First, the low exposure time renders the captured frame useless, which subsequently reduces the frame rate.
Second, in environments with high ambient light levels, such as outdoors on a sunny day, even the shortest exposure time might generate signal charge.
Finally, depending on the image sensor, the noise floor caused by dark current noise could be sufficient to trigger the detection mechanism.

%% file: 09-Conclusion/conclusion.tex
\section{Conclusion}

We have shown that CCD image sensors can be susceptible to intentional electromagnetic interference. Our experiments suggest that this susceptibility stems from the fundamental architecture of CCD image sensors; as the phenomenon remains present across individual designs and yet is absent in devices built on CMOS architectures. 
The presented attack allows an adversary to manipulate the captured frames down to the granularity of single pixels. 
While CCD image sensors are no longer the dominant architecture, they are still widely used in a range of professional applications. 
Therefore, we conclude that the signal injection attacks we have shown pose a serious threat to applications relying on the input from cameras equipped with CCD image sensors. 

%% file: 10-Acknowledgement/acknowledgement.tex
\section{Acknowledgements}

Sebastian Köhler was supported by the EPSRC and the Hans Böckler Foundation.

%% file: 11-Appendix/appendix.tex
\appendix
\renewcommand{\thesection}{\Alph{section}.\arabic{section}}
\setcounter{section}{0}
\nobalance

\begin{appendices}

\begin{figure}[t]
	\centering
	\subcaptionbox{$f_c = 276$~MHz\label{fig:noise_1}}{%
      \includegraphics[width=0.4\textwidth]{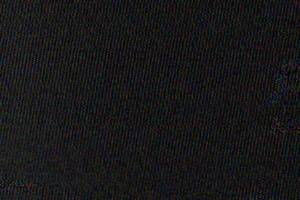}%
      }\par\medskip
    \subcaptionbox{$f_c = 290$~MHz\label{fig:noise_2}}{%
      \includegraphics[width=0.4\textwidth]{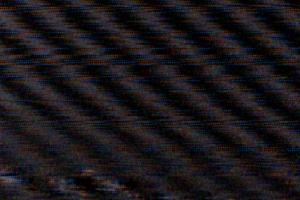}%
      }      
	
	\caption{The same data modulated onto carrier waves with different frequency $f_c$, causes different structured noise.}
	\label{fig:noise_structure}
\end{figure}

\begin{figure*}[t]
	\centering
	\begin{subfigure}[b]{.499\linewidth}
		\centering
		\includegraphics[width=.9\textwidth]{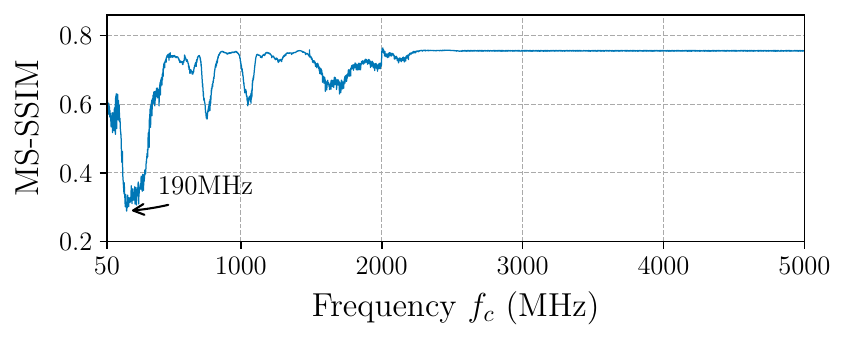}
		\caption{DFM 25G445-ML}
		\label{fig:ms_ssim_25G445} 
	\end{subfigure}%
	\begin{subfigure}[b]{.499\linewidth}
		\centering
		\includegraphics[width=.9\textwidth]{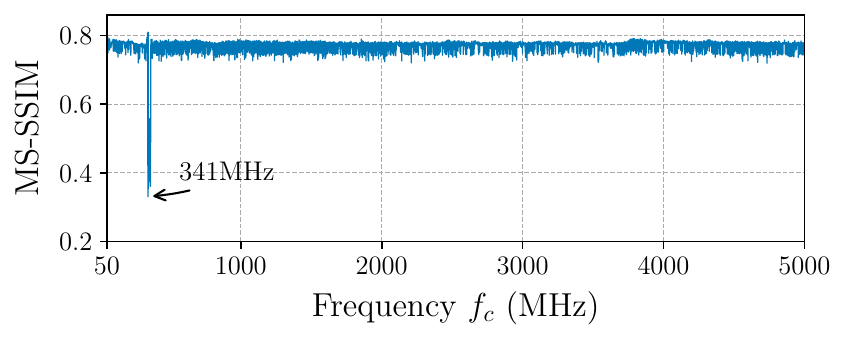}
		\caption{Analog CCD}
		\label{fig:ms_ssim_vhs} 
	\end{subfigure}%
	\caption{Results of the frequency sweep represented in the form of the Multiscale Structural Similarity Index Measure (MS-SSIM) between the frames captured during normal operation and while an attack signal at the carrier frequency $f\textsubscript{c}$ was emitted.}
	\label{fig:ms_ssim_results}
\end{figure*}

\begin{figure*}[t]
	\centering
	\begin{subfigure}[b]{.499\linewidth}
		\centering
		\includegraphics[width=.9\textwidth]{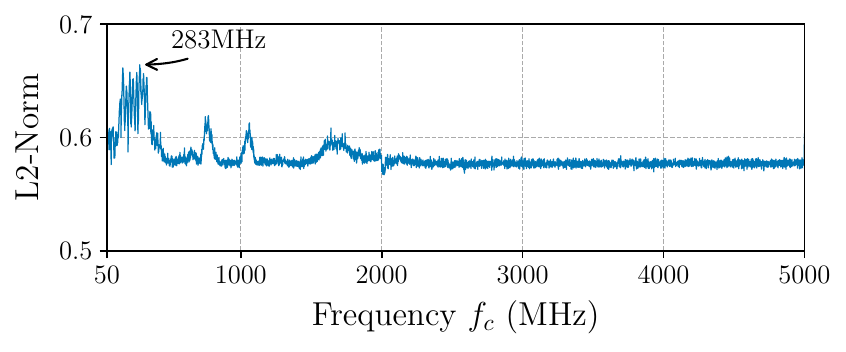}
		\caption{DFM 25G445-ML}
		\label{fig:l2_25G445} 
	\end{subfigure}%
	\begin{subfigure}[b]{.499\linewidth}
		\centering
		\includegraphics[width=.9\textwidth]{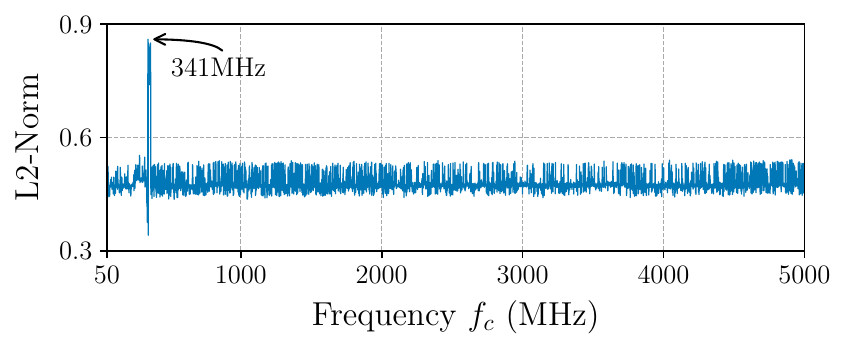}
		\caption{Analog CCD}
		\label{fig:l2_vhs} 
	\end{subfigure}%
	\caption{Results of the frequency sweep represented in the form of the L2-Norm between the frames captured during normal operation and while an attack signal at the carrier frequency $f\textsubscript{c}$ was emitted.}
	\label{fig:l2_results}
\end{figure*}

\begin{figure*}[t]
	\centering
	\begin{subfigure}[b]{.499\linewidth}
		\centering
		\includegraphics[width=.9\textwidth]{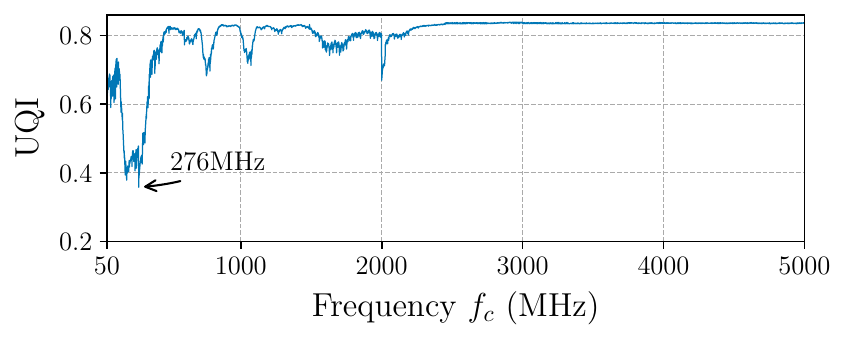}
		\caption{DFM 25G445-ML}
		\label{fig:uqi_25G445} 
	\end{subfigure}%
	\begin{subfigure}[b]{.499\linewidth}
		\centering
		\includegraphics[width=.9\textwidth]{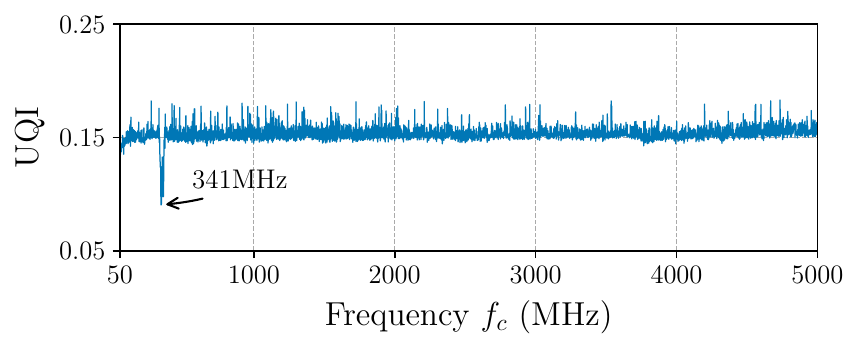}
		\caption{Analog CCD}
		\label{fig:uqi_vhs} 
	\end{subfigure}%
	\caption{Results of the frequency sweep represented in the form of the Universal Image Quanlity Index (UQI) between the frames captured during normal operation and while an attack signal at the carrier frequency $f\textsubscript{c}$ was emitted.}
	\label{fig:uqi_results}
\end{figure*}

\section{Additional Results of the Frequency Sweep to find $f_c$}\label{apx:additional_results}

In this section, we present additional image quality metrics, calculated between legitimate and malicious frames collected during the frequency sweep described in Section~\ref{sec:carrier_freq}.

\paragraph{MS-SSIM}

In addition to the single-scale SSIM image quality metric used throughout the paper, a multi-scale approach, the so-called Multiscale Structural Similarity Index Measure (MS-SSIM), exists~\cite{wang2003multiscale}.
Depending on the parameters of the images that are compared, for instance, the resolution, MS-SSIM performs similar or better than SSIM.
In our case MS-SSIM performed almost identical to single-scale SSIM.
As a result, both metrics determined the same carrier frequency, i.e., $f_c = 190$~MHz, to be the most effective one.

\paragraph{L2-Norm}

The L2-Norm is a commonly used metric in the area of computer vision to highlight discrepancies in semantic information between images~\cite{sinha2011perceptually}.
In Figure~\ref{fig:l2_results}, the results of the frequency sweep for the L2-Norm are presented.
It is immediately recognizable that, in contrast to other image metrics, the most effective carrier frequency is not represented by the smallest value, but rather by the largest.
This is due to the fact that the L2-Norm evaluates the structural properties of an image.
While Gaussian White Noise is random and does not have structured information, the induced noise tends to be more structured.
Surprisingly, using the L2-Norm to determine the most effective carrier frequency $f_c$ for the DFM 25G445-ML leads to a different result compared to SSIM and MS-SSIM.
In contrast, for the analog CCTV board camera, $f_c$ stays the same. 
This is explained by the circumstance that the carrier frequency also influences the structure of the injected noise.
In Figure~\ref{fig:noise_structure}, this difference in noise structure is depicted.
While an attack signal emitted at $f_c = 276$~MHz caused the noise to appear as fine-grained thin bars (Figure~\ref{fig:noise_1}), the same signal appeared as thick bars at $f_c = 290$~MHz (Figure~\ref{fig:noise_2}).

\paragraph{UQI}

The Universal Image Quality Index (UQI) is another metric to measure the quality of an image~\cite{wang2002universal}.
The results of the frequency sweep in form of the UQI are presented in Figure~\ref{fig:uqi_results}.
Similar to the L2-Norm, the most effective carrier frequency for the DFM 25G445-ML is different from the one we observed for SSIM and MS-SSIM.
However, again $f_c$ stays the same for the analog CCTV board camera.

\section{Postprocessed Figures}\label{apx:additional_figures}

In the following, we present Figures~\ref{fig:over_under_sampled} and~\ref{fig:signal_injection_example} after being postprocessed manually with the image editing software GIMP to improve the visibility of the injected distortions for the printed version of this paper. 
More specifically, we increased the brightness, saturation and contrast of the image.

\begin{figure}[H]
	\centering
	\begin{subfigure}[b]{.499\linewidth}
		\centering
		\includegraphics[width=.95\textwidth]{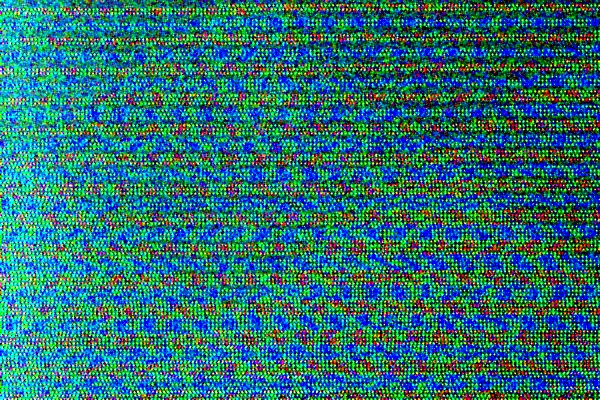}
		\caption{Frame $n-1$}
		\label{fig:frame_1_postprocessed} 
	\end{subfigure}%
	\begin{subfigure}[b]{.499\linewidth}
		\centering
		\includegraphics[width=.95\textwidth]{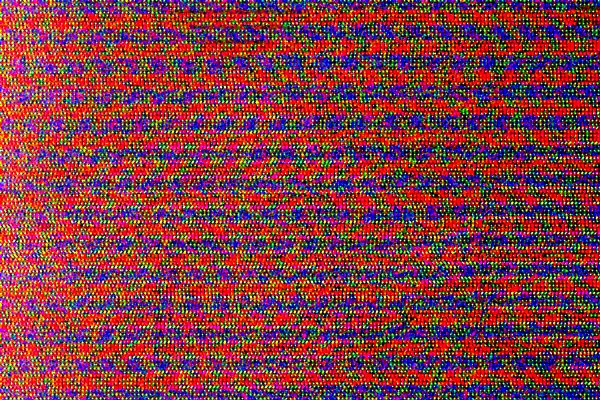}
		\caption{Frame $n$}
		\label{fig:frame_2_postprocessed} 
	\end{subfigure}%
	\caption{This is a postprocessed version of Figure~\ref{fig:over_under_sampled}. Two consecutive example frames captured by the DFM 25G445-ML during the emission of random noise. Due to the wrong sample rate, the injected noise is drifting, causing the stimulus of different photodiodes. As a results, the color of the injected noise changes between consecutive frames.}
	\label{fig:over_under_sampled_postprocessed}
\end{figure}

\begin{figure}[H]
	\centering
	\includegraphics[width=1.0\linewidth]{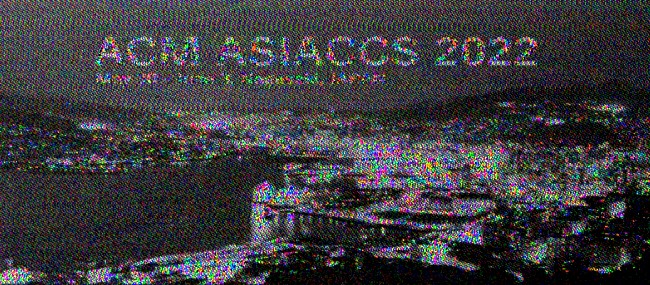}
	\caption{This is a postprocessed version of Figure~\ref{fig:signal_injection_example}. Example of a signal injection attack, illustrating the fine-grained control an adversary can gain over the captured frame.}
	\label{fig:signal_injection_example_postprocessed} 
\end{figure}

\begin{figure}[H]
	\centering
	\includegraphics[width=1.0\linewidth]{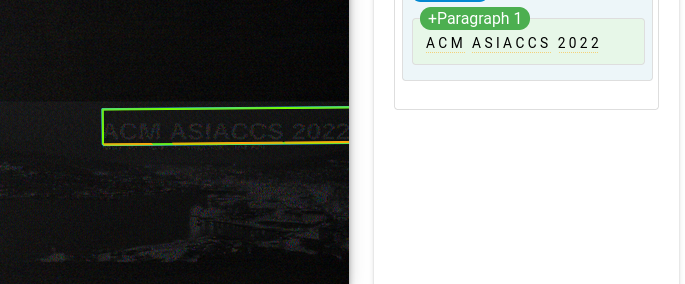}
	\caption{Screenshot of the results from the Google Cloud Vision API. The injected text was correctly recognized.}
	\label{fig:google_api_result} 
\end{figure}

\end{appendices}